\documentclass[pre,twocolumn,superscriptaddress,preprintnumbers,amsmath,amssymb,footinbib]{revtex4}
\usepackage[dvips]{graphicx}
\usepackage{dcolumn}
\usepackage{color}
\usepackage{bm}

\usepackage{epstopdf}

\begin{document}

\def\bra#1{\mathinner{\langle{#1}|}}
\def\ket#1{\mathinner{|{#1}\rangle}}
\def\braket#1{\mathinner{\langle{#1}\rangle}}
\def\Bra#1{\left\langle#1\right|}
\def\Ket#1{\left|#1\right\rangle}

\title{Quantum jump model for a  system with a finite-size environment}

\author{S. Suomela}
\affiliation{Department of Applied Physics and COMP Centre of Excellence,
Aalto University School of Science, P.O. Box 11100, 00076 Aalto, Finland}

\author{A. Kutvonen}
\affiliation{Department of Applied Physics and COMP Centre of Excellence,
Aalto University School of Science, P.O. Box 11100, 00076 Aalto, Finland}

\author{T. Ala-Nissila}
\affiliation{Department of Applied Physics and COMP Centre of Excellence,
Aalto University School of Science, P.O. Box 11100, 00076 Aalto, Finland}
\affiliation{Department of Physics, P.O. Box 1843, Brown University, Providence, Rhode Island 02912-1843, U.S.A.}

\begin{abstract}
Measuring the thermodynamic properties of open quantum systems poses a major challenge. A
calorimetric detection has been proposed as a feasible experimental scheme to measure work 
and fluctuation relations in open quantum systems. However, the detection requires a finite size for the environment, which influences the system dynamics. This process cannot be modeled with the standard stochastic approaches. We develop a quantum jump model suitable for systems coupled to a finite-size environment. With the method we study the common fluctuation relations and prove that they are satisfied.

\end{abstract}

\date{January 13, 2015}

\maketitle

\section{Introduction}
Rapid progress in the fabrication and manipulation of micro and nanoscale objects \cite{nakamura2010nonequilibrium,kung2012irreversibility,pekola2015towards,rossnagel2015single} has made it necessary to extend the concepts of thermodynamics to small systems which by definition are not in the thermodynamic limit. 
In such systems the extensive thermodynamic quantities, such as entropy, 
heat and work, are not described by their average values
alone but due to fluctuations they obey nontrivial probability distributions. Fortunately, it has been shown that in many cases the stochastic thermodynamic variables obey fluctuation relations  \cite{spjetp45/125,prl78/2690,Seifert2005} which often appear in the form of relations between exponential averages of the extensive variables. 

While the two-measurement protocol of thermodynamic variables, especially work,  is now well studied in closed quantum systems, there have been conceptual problems in open quantum systems
\cite{pre73/046129, prl102/210401, PhysRevLett.107.140404,Crooks2007, jsp148/480, pre88/032146, rmp81/1665,rmp83/771, pre89/012127, pre89/032114, PhysRevE.90.022103, 1367-2630-16-11-115001, PhysRevE.89.052128, PhysRevB.90.075421, Jarzynski2015quantum,PhysRevLett.113.140601,an2014experimental,Deffner2015,carrega2015, PhysRevB.91.224303,1367-2630-17-7-075018,PhysRevE.85.031110, prl111/093602,NJP15/085028,PhysRevA.88.042111, PhysRevE.89.042122, suomela2014moments,s10955-014-0991-1,liu2015calculating,PhysRevE.91.062109,Suomela2015a,PhysRevE.92.032129}. To make connection to classical stochastic thermodynamics, the quantum jump (QJ) method \cite{prl68/580, pra45/4879, pra46/4363, OSAQO, rmp70/101} has been recently used to study thermodynamics and fluctuation theorems in open quantum systems as it tries to mimic the trajectories realized in actual experiments 
\cite{PhysRevE.85.031110,prl111/093602, NJP15/085028,PhysRevA.88.042111,PhysRevE.89.042122,suomela2014moments,Suomela2015a,liu2015calculating,s10955-014-0991-1,PhysRevE.91.062109,PhysRevE.92.032129}. The method unravels the master equation of the reduced density matrix as stochastic trajectories with environment induced jumps between the system states. 
The concepts of stochastic thermodynamics can be developed by associating a jump with heat exchange. However, there are several approximations that limit the generality of the QJ method. In particular, it is only applicable assuming a memoryless or an infinitely large environment, i.e. an ideal heat bath whose state
does not change during the drive.

Even within the QJ framework the issue of actually measuring the energy change in a driven open quantum
system is nontrivial. It has been proposed by J. P. Pekola et al. \cite{Pekola2012} that this could be done by a so-called calorimetric measurement, where the immediate environment itself measures the energy change in the system. In Fig. 1, we show a schematic of such a setup in the case where there is a driven qubit coupled to a finite-size calorimeter and an ideal heat bath. The key point in the calorimetric measurement is that in order to observe the energy changes of the system, the calorimeter must be finite, i.e., in contrast to the ideal bath it has to change its state when absorbing energy from the system \cite{Pekola2012, gasparinetti2015fast, Suomela2015b}.

 \begin{figure}[t]
    \begin{center}
    \includegraphics[scale=.35]{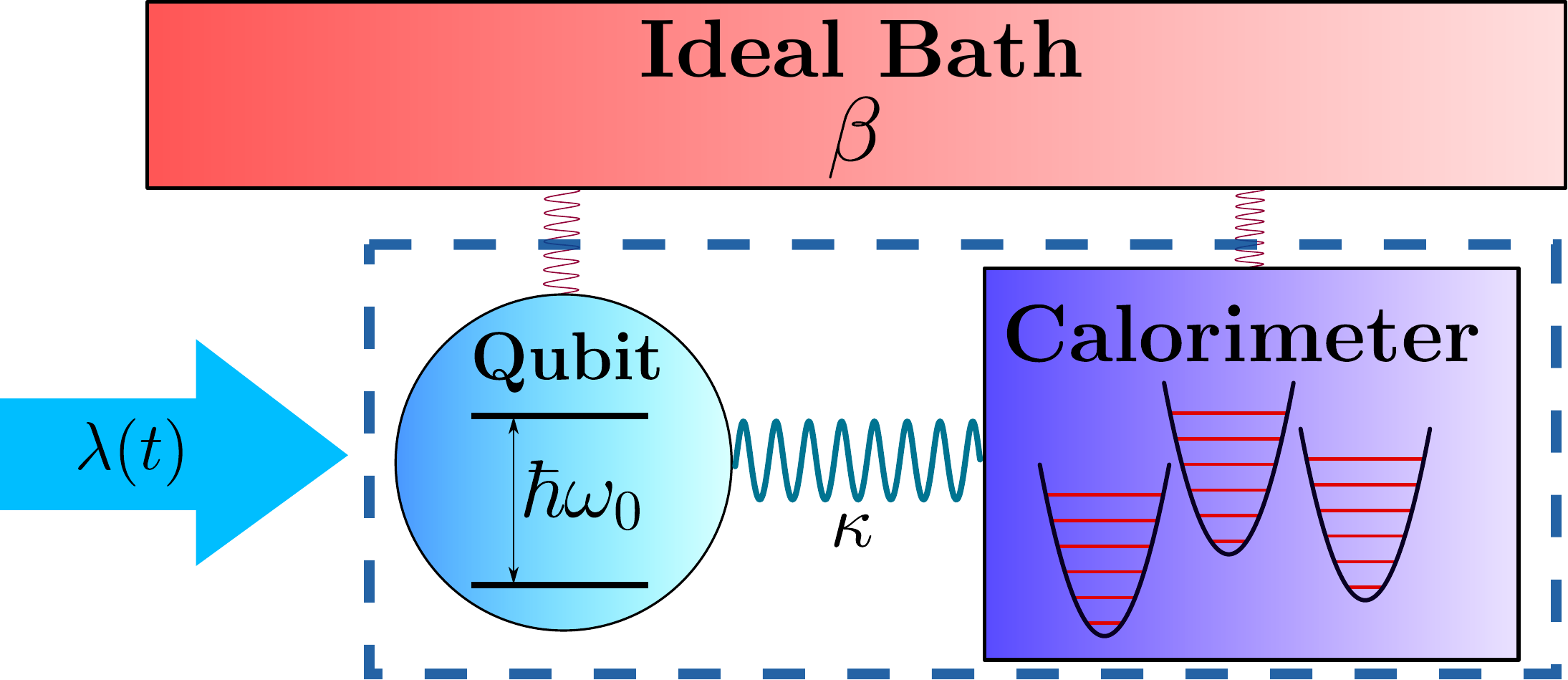}
    \end{center}
    \caption{A schematic of the calorimetric measurement in a driven open quantum system. Here, a qubit is coupled to the calorimeter that is described by finite number of harmonic oscillators with an energy gap equivalent to that of the qubit. The qubit and the calorimeter are initially thermalized with an ideal bath of inverse temperature $\beta$. The qubit is driven by a classical source $\lambda(t)$. }
    \label{fig:syst}
\end{figure}

The standard QJ approach is not applicable to analysis of the calorimetric measurement setup, since the calorimeter is not ideal bath. In this letter, we develop a modified QJ model suitable for systems weakly coupled to a finite-size environment, called calorimeter from here on. In the model, a jump in the system changes both the state of the system and the state of the calorimeter. Due to the influence of the system on the calorimeter's evolution, the system evolution is no longer Markovian as the previous history of the system affects its future evolution via the state of the calorimeter. With the new method, we show that the common fluctuation relations are satisfied
for the system-calorimeter composite. As a concrete example, we numerically study a sinusoidally driven qubit weakly coupled to the calorimeter which comprises harmonic oscillators with an energy gap equivalent to that of the qubit. The qubit and the calorimeter are initially thermalized with an ideal bath of inverse temperature $\beta$ as depicted in  Fig. \ref{fig:syst}.

\section{The standard quantum jump method}
Let us first give a short introduction to the standard QJ method in the literature
\cite{prl68/580, pra45/4879, pra46/4363, OSAQO, rmp70/101}. Instead of evolving the density matrix as done in the direct master equation calculations, the quantum jump method unravels the master equation into quantum trajectories with stochastically evolving wave functions. For a single trajectory, the probability for a jump between time $t$ and $t+\delta t $ is given by
\begin{eqnarray}
\delta p =  \sum_m \delta p_m = \sum_m \delta t \bra{\psi(t)} \hat C^\dagger_m  \hat C_m  \ket{\psi(t)}, 
\end{eqnarray}
for $\delta t \rightarrow 0$, where $\ket{\psi(t)}$ is the state of the system before the jump, $\delta p_m$ is the probability for a jump corresponding to a jump operator $\hat C_m=\sqrt{\Gamma_m} \hat A_m$. The matrix $\hat A_m$ gives the form of the jump and $\Gamma_m$ is the transition rate. If  a jump corresponding to $\hat C_m$ occurs, the new  state is given by $\ket{\psi(t+\delta t)}=\hat C_m  \ket{\psi(t)} / \sqrt{(\delta p_m/\delta t)}$, where $\delta p_m/\delta t$ normalizes it. If no jumps occur during the time interval $[t,t+\delta t]$, the state evolution is not given by the system Hamiltonian $\hat H_s(t)$ alone but by the non-unitary Hamiltonian
$
\hat H(t)= \hat H_s(t)-\frac{i \hbar}{2} \sum_m \hat C^\dagger_m  \hat C_m,
$
yielding $ \ket{\psi(t+\delta t)}= \frac{1}{\sqrt{1-\delta p}} \left(1-\frac{i}{\hbar} {\hat H}(t) \delta t \right)  \ket{\psi(t)}$. Although the jump operator can be time-dependent, the past history of the trajectory, e.g., the number of jumps, does not affect the jump operators at all. As a consequence, the evolution of a stochastic trajectory depends only on the current state of the system $\ket{\psi(t)}$. This is a good approximation when the environment is ideal.  However, such an environment makes the calorimetric measurement infeasible as the system evolution leaves no traces to the environment.

\section{Quantum jump method with a finite environment}
We now wish to extend the standard QJ method to the case corresponding to Fig. 1, where transitions in the system influence the state of
the calorimeter, both initially thermalized with an ideal heat bath.  We assume the calorimeter to be large enough to allow a semiclassical treatment such that there is an orthonormal eigenbasis where the calorimeter density matrix is diagonal. In practice, this basis will be the einselected basis of the calorimeter determined by the detector and the super bath \cite{Zurek2003}. We call these eigenbasis states as microstates. We also assume the system-calorimeter coupling to be weak enough such that it can be neglected in the energy terms and modeled by stochastic jumps alone. We take into account only transitions that conserve the energy of the calorimeter-system composite.  

To be precise, instead of using jump operators that only depend on the system degrees of freedom, 
we define new jump operators $\hat D_m=g_m  \hat A_m \otimes \hat B_m$, where $ \hat A_m$ causes a transition between system states and $\hat B_m$ between the calorimeter microstates defined above \cite{Note1}. The coefficient $g_m$ is proportional to the coupling strength. We define the probability for a transition in the time interval $[t,t+\delta t] $ as
\begin{eqnarray}
\delta p =  \sum_m \delta p_m = \sum_m \delta t \text{Tr}_{s+c} \lbrace \hat D^\dagger_m  \hat D_m \hat \sigma_s \otimes \hat \sigma_c \rbrace, \label{eq:dp1}
\end{eqnarray}
where $\delta p_m$ is the probability for a jump corresponding to the jump operator $\hat D_m$,  
$\hat \sigma_s(t)=\ket{\psi(t)} \bra{\psi(t)}$ is the matrix form of the system state, and  
$\hat \sigma_c(t)=\ket{\Psi(t)} \bra{\Psi(t)}$ is the instantaneous calorimeter microstate,
and the trace is over both the system ($s$) and calorimeter ($c$) degrees of freedom. 
If a jump corresponding to $\hat D_m$ occurs, the new system and calorimeter states are given by $\hat \sigma_s(t+\delta t)= {\rm Tr}_c \lbrace \hat D_m  \hat \sigma_s \otimes \hat \sigma_c \hat D_m^\dagger \rbrace /  (\delta p_m / \delta t)$ and  $\hat \sigma_c(t+\delta t)= {\rm Tr}_s \lbrace \hat D_m  \hat \sigma_s \otimes \hat \sigma_c \hat D_m^\dagger \rbrace/ (\delta p_m / \delta t) $. 

If no jumps occur during the time interval $[t,t+\delta t]$, the time evolution is given by the non-unitary Hamiltonian
\begin{eqnarray}
\hat H(t)= \hat H_s(t)+ \hat H_c-\frac{i \hbar}{2} \sum_m \hat D^\dagger_m  \hat D_m,\label{eq:dp2}
\end{eqnarray}
where $\hat H_s$ and $\hat H_c$ are the system and calorimeter Hamiltonians, respectively. The new system state 
$\hat \sigma_s(t+\delta t)= {\rm Tr}_c \lbrace \hat U(t+\delta t,t) \hat \sigma_s \otimes \hat \sigma_c \hat U^\dagger(t+\delta t,t) \rbrace / (1- \delta p) +\mathcal{O}(dt^2)$, where $\hat U(t+\delta t,t)= 1- \frac{i}{\hbar}  \hat{H}(t) \delta t$. Similarly, the new calorimeter 
state  $\hat \sigma_c(t+ \delta t)= {\rm Tr}_s \lbrace \hat U(t+\delta t,t) \hat \sigma_s \otimes \hat \sigma_c \hat U^\dagger(t+\delta t,t) \rbrace/(1-\delta p) +\mathcal{O}(dt^2)$, which gives $\hat \sigma_c(t+ \delta t)= \hat \sigma_c(t)$ as we assumed that $\hat \sigma_c(t)$ is in a microstate. As a consequence, it is sufficient to focus in detail only on the system dynamics, 
where the calorimeter's state only affects the transition rates.

\section{Fluctuation relations}
For studying stochastic thermodynamics and the associated fluctuation relations with the method, we focus on a generic two-level system (qubit) with 
$\hat H_0 = \hbar\omega_0 \hat a^\dagger \hat a$ that is weakly driven by a classical source $V(t)=\lambda(t)(\hat a+ \hat a^\dagger)$, where $\hat{a}$ and $\hat{a}^\dagger$ are the annihilation and creation operators in the undriven basis. The system Hamiltonian is then given by $\hat{H}_s(t)=\hat{H}_0+\hat{V}(t)$. The qubit is coupled to the calorimeter by $\hat V_N(t)=\kappa \sum_m (\hat{a}^\dagger \hat{b}_m  + \hat{a} \hat{b}_m^\dagger)$, where the coupling strength $\kappa$ is real and the operators $\hat b_m$  depend on the exact form of the calorimeter. The calorimeter Hamiltonian is given by $\hat H_c =\sum_n \epsilon_n \hat d_n^\dagger \hat d_n$. For a bosonic calorimeter $\hat d_n$ and $\hat d^\dagger_n$ are the annihilation and creation operators associated with energy $\epsilon_n$ and the operators $\hat b_m$ form a set of all the annihilation operators associated with energy $\hbar \omega_0$, i.e,  $\sum_m \hat b_m =\sum_n \hat d_n \delta_{\epsilon_n,\hbar \omega_0}$ \cite{Note3}. Before and after the driving protocol, both the qubit and the calorimeter states are measured by monitoring the calorimeter state only. The qubit state can be indirectly determined from the previous jump before the drive and from the next jump after the drive in the calorimeter. However, this calorimetric monitoring is equivalent to performing the measurements using the two measurement protocol for both the qubit and the calorimeter, as shown in Appendix A. 

The total system is initially prepared such that both the qubit and the calorimeter start as a pure state given by the joint probability $P[i,\Psi_0]$, where $\ket{i}$ and $\ket{\Psi_0}$  are the initial qubit and calorimeter states, respectively. Similar to the standard perturbative treatment \cite{Girvin2010}, we define the jump operators in the undriven basis, i.e., $\hat D_{\uparrow,m}=g_m  \hat a^\dagger \otimes \hat b_m$ and  $\hat D_{\downarrow,m} =g_m \hat a \otimes \hat b_m^\dagger$, where the coefficient $g_m \propto \kappa$.  Due to the energy change with the qubit, the calorimeter can jump to a different microstate such that the energy difference of the microstates corresponds to the energy change in the qubit. If the calorimeter is assumed to stay in the same microstate between the jumps, we can write the calorimeter traced jump operators \cite{Note2} as $\hat C_{\uparrow,m}(\Psi_k)=\sqrt{\Gamma_{\uparrow,m}(\Psi_k)} \hat a^\dagger$ and $\hat C_{\downarrow,m}(\Psi_k)=\sqrt{\Gamma_{\downarrow,m}(\Psi_k)} \hat a$, where the transition rates are
\begin{eqnarray}
\Gamma_{\uparrow,m}(\Psi_k)&=& |g_k|^2  {\rm Tr}_c\lbrace \hat b_m \hat \sigma_c(\Psi_k) \hat b_m^\dagger \rbrace, \\
 \Gamma_{\downarrow,m}(\Psi_k)&=& |g_k|^2  {\rm Tr}_c\lbrace \hat b_m^\dagger \hat \sigma_c(\Psi_k) \hat b_m \rbrace,
 \end{eqnarray}
with $\hat \sigma_c(\Psi_k)=\ket{\Psi_k}\bra{\Psi_k}$. If a jump caused by $\hat C_{\downarrow,m}(\Psi_k)$ occurs, the new calorimeter state is $\ket{\Psi_{k+1}}=\hat b_m^\dagger \ket{\Psi_k}/ || \hat b_m^\dagger \ket{\Psi_k}||$. According to stochastic thermodynamics \cite{crooks1999entropy,Parrondo2015}, the entropy production associated with a jump is defined as the logarithmic ratio of the forward and backward transition rates. Due to the symmetry $\Gamma_{\downarrow,m}(\Psi_k)=\Gamma_{\uparrow,m}(\Psi_{k+1})$, this entropy production is always zero and the total entropy production of a trajectory depends only on the initial and final states of the qubit  and the calorimeter, i.e., 
\begin{equation}
\Delta S_{T} =  \ln \left\lbrace  {P[i,\Psi_0]}/{\bar{P}[f,\Psi_N]}\right\rbrace,
\end{equation} 
where $\bar{P}[f,\Psi_N]$ is the probability to start a reversed trajectory with the final qubit state $\ket{f}$ and the final calorimeter state $\ket{\Psi_N}$ of the forward trajectory. The entropy production satisfies the fluctuation theorem (see Appendix B for details):
\begin{equation}
\langle e^{-\Delta S_{T}} \rangle  = 1, \label{eq:stot} 
\end{equation} 
where the average is over all the forward trajectories. If the initial probability distribution of the forward trajectory follows the 
canonical ensemble in equilibrium with the ideal heat bath, the probability distribution of the reversed trajectories can be chosen to follow a canonical ensemble with the same temperature. By defining the work $W$ associated with a single trajectory as the energy difference between the final and initial states of the qubit-calorimeter composite, Eq. \eqref{eq:stot} gives the Jarzynski equality for work as
\begin{equation}
\langle e^{-\beta W } \rangle = e^{-\beta \Delta F}, \label{eq:JE} 
\end{equation}
where $\Delta F$ is the free-energy difference between the final and initial states. 

The results above were derived assuming that the calorimeter stays in the same microstate until the next jump. However, in many systems such as in electronic devices \cite{jsm2013/P02033,Kutvonen2015}, the relaxation rate inside the calorimeter is the fastest time scale. 
Consequently, the calorimeter does not stay in a single microstate between jumps but shifts quickly between the microstates that correspond to the same energy. We can still calculate the qubit dynamics with Eqs. \eqref{eq:dp1} and \eqref{eq:dp2} by using an averaged calorimeter state instead of a single calorimeter microstate. According to the microcanonical ensemble, the averaged calorimeter state %
$
\hat \sigma_c(E)= (1/N(E)) \sum_\Psi \ket{\Psi}\bra{\Psi} \delta_{E_\Psi,E}, 
$
where the sum is over all the microstates, $N(E)$ is the number of microstates with energy $E$ and $E_\Psi$ is the energy of microstate $\ket{\Psi}$. Let us call this state a macrostate. We assume that calorimeter reaches the macrostate instantaneously after a jump. The probability to start with the qubit state $\ket{i}$ and the calorimeter macrostate of energy $E_0$ is given by $P[i,E_0]$.  Due to the energy change with the qubit, the calorimeter can jump to another macrostate. As the calorimeter reaches the macrostate immediately after a jump, we can sum over all the transition rates corresponding the same energy change. The resulting total transition rates are given by
\begin{eqnarray}
\Gamma_{\downarrow}(E) = \frac{1}{N(E)} \sum_{m, \Psi} \Gamma_{\downarrow,m}(\Psi)  \delta_{E_\Psi,E}, \label{ETR1} \\
\Gamma_{\uparrow}(E) =  \frac{1}{N(E)} \sum_{m, \Psi} \Gamma_{\uparrow,m}(\Psi)  \delta_{E_\Psi,E}  \label{ETR2},  \end{eqnarray}
and they satisfy the detailed balance condition 
\begin{equation}
{\Gamma_{\downarrow}(E-\hbar\omega_0) }/{\Gamma_{\uparrow}(E) }={N(E )}/{N(E-\hbar \omega_0)}, \label{eq:DBC2}
\end{equation}
which resembles the fluctuation relation derived for microcanonical ensembles \cite{PhysRevLett.96.050601,PhysRevE.88.042136}.  
The entropy production of a jump up with the calorimeter energy $E$ is given by
\begin{eqnarray}
\Delta s_\uparrow(E) = -  \ln \left( \frac{\Gamma_{\downarrow}(E-\hbar\omega_0) }{\Gamma_{\uparrow}(E )} \right) = \ln \left( \frac{N(E-\hbar \omega_0)}{N(E)} \right), \label{DBCE2}
\end{eqnarray}
which gives a natural interpretation of the entropy production as the Boltzmann entropy change of the calorimeter. The entropy productions of up and down jumps are related by $\Delta s_\uparrow(E)=-\Delta s_\downarrow(E-\hbar \omega_0)$. The total entropy production is 
then given by
\begin{eqnarray}
\Delta S_{T}=  \ln \left\lbrace  {P[i,E_0]}/{\bar{P}[f,E_N]}\right\rbrace+\sum_{i=1}^N \Delta s_{\chi_i}{(E_{i-1})},
\end{eqnarray}
where $N$ is the number of jumps, $E_i$ is the calorimeter energy after the $i^{\rm th}$ jump,  $\chi_i=\uparrow/\downarrow$ is the 
direction of $i^{\rm th}$ jump, and $\bar{P}[f,E_N]$ is the probability to start a reversed trajectory with the forward trajectory's final qubit state $\ket{f}$ and calorimeter energy $E_N$. As Eq. \eqref{eq:stot} still holds,  we recover the Jarzynski equality if we start from the canonical ensemble.

\section{Numerical results}
To illustrate the method, we have done numerical simulations of the coupled qubit-calorimeter composite, where the calorimeter is
described by $n$ quantum harmonic oscillators, with energy gap equivalent to that of the qubit $\hbar \omega_0$. 
The harmonic oscillators can be non-interacting or they interact fast enough such that the calorimeter can be assumed to reach the macrostate instantenously after a jump. In both cases, the transition rates are the 
same and depend only on the calorimeter energy:
\begin{eqnarray}
\Gamma_{\downarrow}(E) = |g|^2 (E+n\hbar \omega_0), \quad\Gamma_{\uparrow}(E) =  |g|^2 E,\label{ETR2} 
\end{eqnarray}
where we have for convenience chosen $E=0$ when all the oscillators are in the ground state. Consequently, the qubit's evolution depends only on the energy of the calorimeter.  As the calorimeter energy does not change between the jumps, we can calculate the qubit dynamics using Eqs. \eqref{eq:dp1} and \eqref{eq:dp2} with environment traced jump operators $\hat{C}_\downarrow=\sqrt{\Gamma_{\downarrow}(E)}\hat a$ and $\hat{C}_\uparrow=\sqrt{\Gamma_{\uparrow}(E)}\hat a^\dagger$ \cite{Note2}. The calculations are done in the interaction picture with respect to $\hat H_0+\hat H_e$.

In order to compare the results with different number of oscillators, we use the value $|g|^2=0.025 / ( n \hbar )$ in the simulations such that the total coupling strength remains the same. We start the qubit and the calorimeter from a canonical ensemble with respect to the inverse temperature $\beta = 1/(\hbar \omega_0)$. The qubit is driven sinusoidally with $\lambda(t)=0.05\hbar \omega_0 \sin (\omega_0 t)$. The protocol consist of driving, no driving, driving, no driving parts each lasting a time interval equal of 50 driving periods. %
\begin{figure}[t!]
    \begin{center}
    \includegraphics[scale=.4]{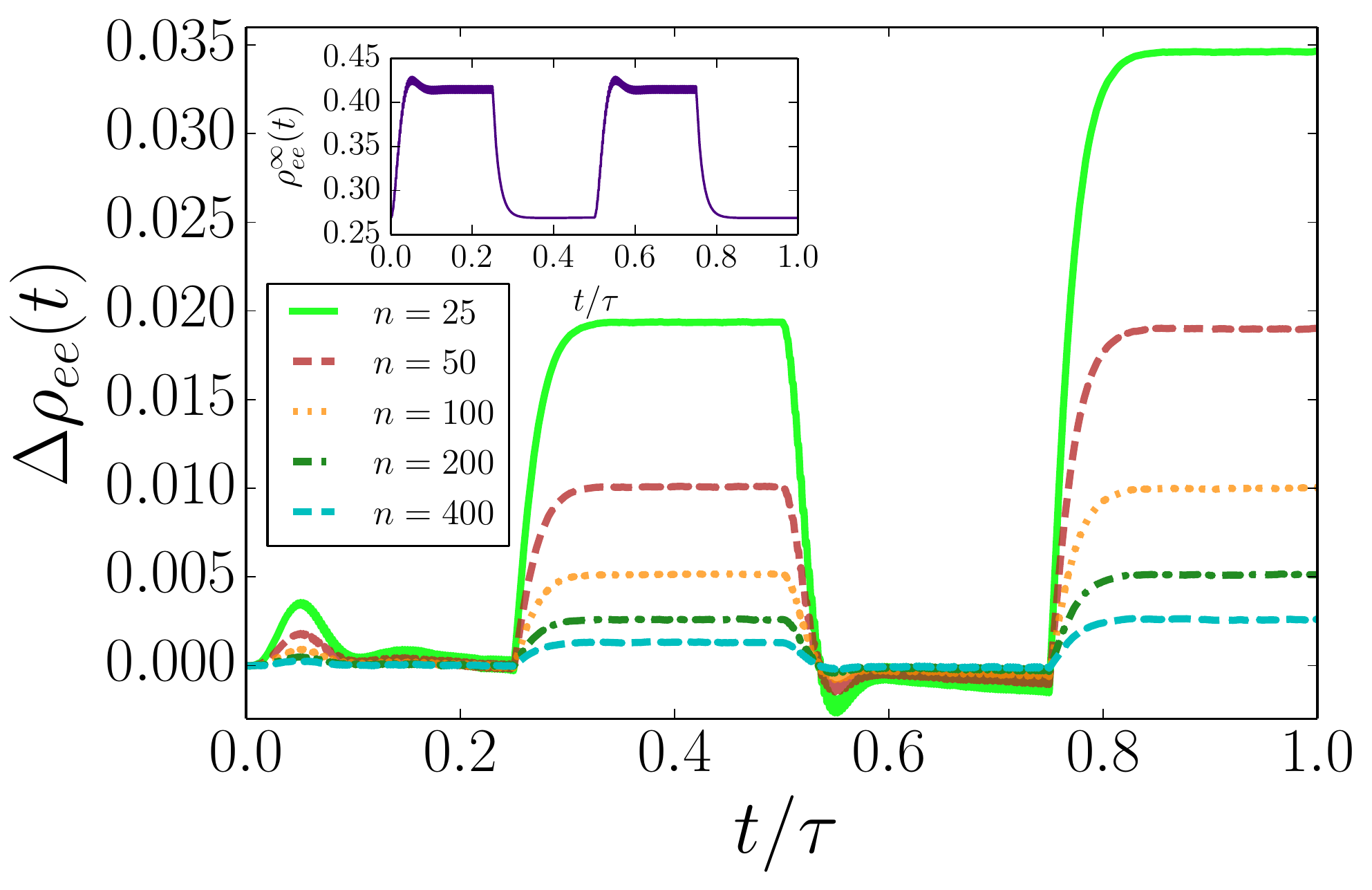}
    \end{center}
    \caption{The difference between the populations of the qubit excited state for $n$ and infinite number of oscillators 
    $\rho_e(t) \equiv \rho^n_e(t) - \rho^\infty_e(t)$. The data is from numerical simulations of the coupled qubit-calorimeter system. The protocol ends at time $\tau$. The inset shows the excited state population when $n=\infty$. The drive is discretized using $2 \times 10^5$ time steps and the number of trajectories is $10^9$. See text for the values of the other system parameters.}
    \label{fig:pop}
\end{figure}

\begin{figure}[t!]
    \begin{center}
    \includegraphics[scale=.4]{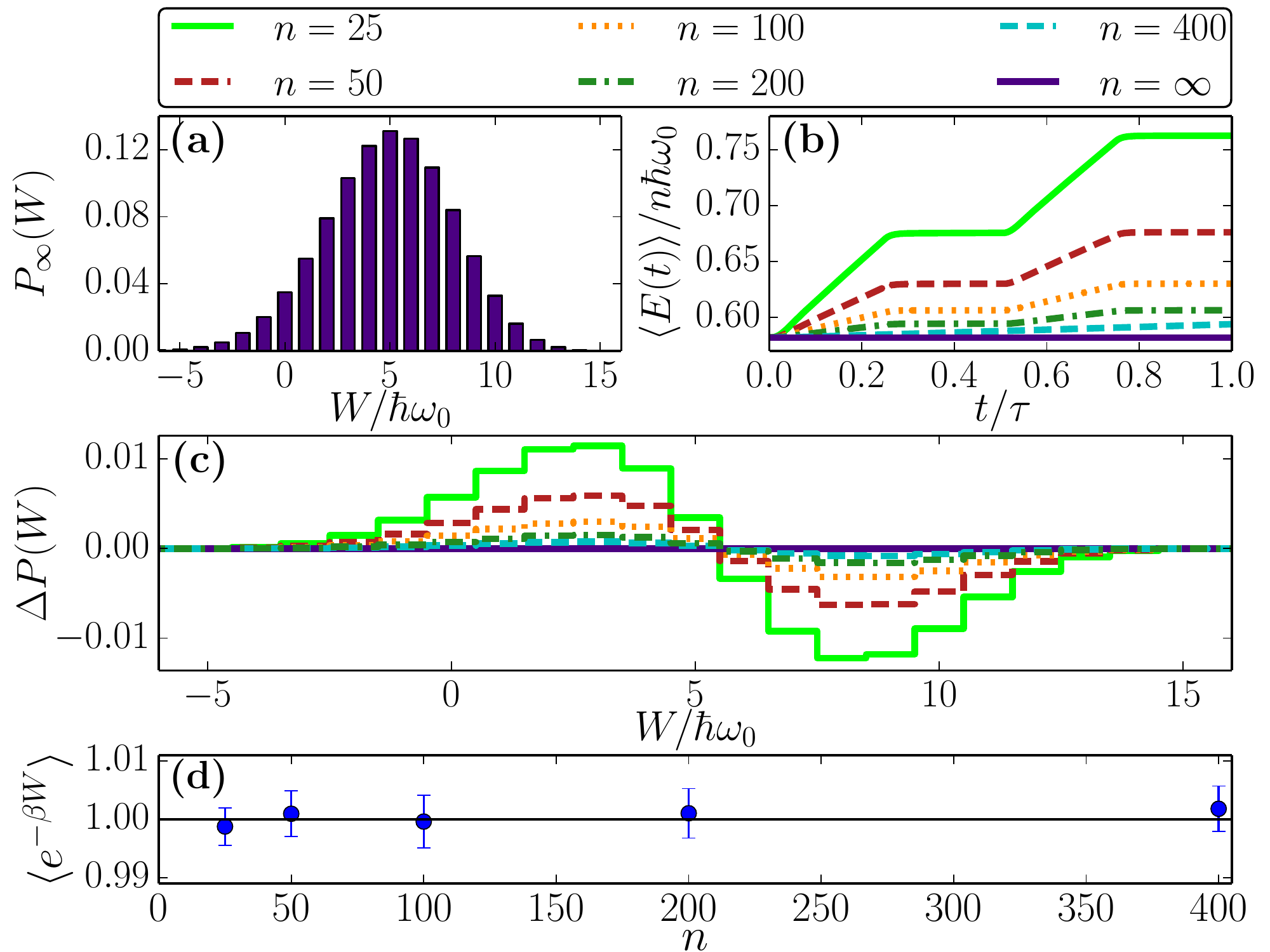}
    \end{center}
    \caption{Influence of finite-size calorimeter on the work and energy statistics when it consists of $n$ harmonic oscillators.  (a)  The probability distribution of work in the case of $n=\infty$, $P_\infty(W)$.  (b) Time evolution of the average calorimeter energy for different values
    of $n$. (c) Deviation of the probability distribution of work $\Delta P(W) \equiv P_n(W) - P_\infty(W)$. 
    (d) Numerical verification of the Jarzynski equality. The error bars are the standard deviation times 1.96 and correspond to a 95 \% confidence interval. The parameters are the same as in Fig. \ref{fig:pop}.}
    \label{fig:work}
\end{figure}

Figure \ref{fig:pop} nicely illustrates the influence of the calorimeter's size on the qubit dynamics. Due to its finite size, the calorimeter 
is driven out of equilibrium. During the drive, the effect of the calorimeter on the qubit dynamics is suppressed by the drive 
since it is stronger than the qubit-calorimeter coupling. However, when the drive is stopped and the qubit equilibrates, the effect of the calorimeter becomes apparent. The drive causes the qubit to emit energy to the calorimeter. For a very large calorimeter (here
the largest $n=400$), this additional energy is very small as compared to the initial energy, which scales with size. However, for a smaller $n$ the change in the relative energy becomes more pronounced as depicted in Fig. \ref{fig:work}(b).

For the setup, the work is obtained as the energy difference between the final and initial states of the qubit plus the heat released to the calorimeter. This yields the same result as the two-measurement protocol for the composite. As shown in Figs. \ref{fig:work}(a) and (c), the finite size of the calorimeter causes the work distribution to deviate from the inifinite-size limit as overheating changes the transition rates.  The transition rates are strongly influenced by
the previous jumps for small values of $n$. This is illustrated in Fig. \ref{fig:work}(c), where work values between $\pm 5 \hbar \omega_0$ are more probable for small $n$. Independent of $n$, the work distributions were found to be consistent with the Jarzynski equality within the statistical errors, as shown in Fig. \ref{fig:work}(d).

\section{Conclusions}
Measurement and definition of work in driven open quantum systems poses an interesting theoretical and experimental challenge. 
In the present work, we have considered the proposal of a calorimetric measurement that offers a simple and transparent way of
measuring energy exchange between the system and the calorimeter. To be able to theoretically analyse a calorimetric
setup, we have developed a modified QJ model suitable for systems with a finite-size environment. 
We have shown that  due to the finite size, the calorimeter is driven out of equilibrium leading to changes in the reduced system's 
dynamics and in the work statistics. These changes cannot be modelled with the standard QJ method which assumes that the
whole environment is an ideal heat bath. With the model, we have analytically and numerically shown that the standard fluctuation relations are still valid and remain unaltered by the finite size of the calorimeter.  This is in contrast to Ref. \cite{jsm2013/P02033}, where the transition rates violate the detailed balance condition of Eq. \eqref{eq:DBC2} (see Appendix C).

\acknowledgements
We wish to thank Jukka Pekola for suggesting this problem to us, and for useful discussions. We wish also to thank C. Jarzynski, P. Muratore-Ginanneschi and K. Schwieger for useful discussions and comments. This work was supported in part by the V\"ais\"al\"a foundation and the Academy of Finland through its Centres of Excellence Programme (2015-2017) under project numbers 251748 and 284621. The numerical calculations were performed using computer resources of the Aalto University School of Science "Science-IT" project.

\appendix

\section{Equivalence between the calorimetric measurement and the two measurement protocol} 
 Let us assume that after the drive, the qubit is in a superposition state $\ket{\psi}=\alpha_0 \ket{0}+\alpha_1 \ket{1}$, where $\ket{0}$ and $\ket{1}$  are the ground and  exicted states of the undriven qubit, respectively. The calorimeter is assumed to have energy $E$. A double projection measurement of both the qubit and the calorimeter then gives $E$ with probability $P(E)=|\alpha_0|^2$ and $E+\hbar \omega_0$ with probability $P(E+\hbar \omega_0)=|\alpha_1|^2=1-|\alpha_0|^2$.
 
In the calorimetric measurement, the  qubit state is measured by waiting that the qubit collapses to an eigenstate due to a jump. Let us first study the case where the calorimeter is in a single microstate $\ket{\Psi_k}$ and it does not change between the jumps. In this case, we can use the calorimeter traced jump operators $\hat C_{\uparrow, m}(\Psi_k)=\sqrt{\Gamma_{\uparrow, m}(\Psi_k)} \hat{a}^\dagger$ and $\hat C_{\downarrow, m}(\Psi_k)=\sqrt{\Gamma_{\downarrow, m}(\Psi_k)} \hat{a}$. The probability that the first jump after the drive is caused by jump operator $\hat C_{\downarrow, m}$ takes the form 
\begin{eqnarray}
P_{\downarrow,m}&=&\int_0^\infty dt  \Gamma_{\downarrow,m}(\Psi_k) \times \nonumber \\ &&\left\lvert \bra{1}  e^{-\frac{1}{2}\sum_n [\Gamma_{\downarrow,n}(\Psi_k) \ket{1}\bra{1}+\Gamma_{\uparrow,n}(\Psi_k) \ket{0}\bra{0}]t} \ket{\psi} \right\rvert^2  \nonumber \\
 &=& \int_0^\infty dt {\Gamma_{\downarrow,m} (\Psi_k)} |\alpha_1|^2 e^{-\sum_n \Gamma_{\downarrow,n} (\Psi_k)t} \nonumber \\ &=& \frac{|\alpha_1|^2 \Gamma_{\downarrow,m}(\Psi_k)}{\sum_n \Gamma_{\downarrow,n}(\Psi_k)}.
 \end{eqnarray}
 The probability that the first jump after the drive is a jump down is obtained by summing over all $m$,
 \begin{eqnarray}
P_{\downarrow}&=\sum_m P_{\downarrow,m}= |\alpha_1|^2.  \end{eqnarray}
  In the case of a jump down, the qubit is known to be in the ground state after the jump and thus the total energy measured is simply the calorimeter energy after the jump, $E+\hbar \omega_0$. Similarly, the probability that the first jump after the drive is a jump up takes the form
\begin{eqnarray}
P_\uparrow&=&\sum_m \int_0^\infty dt \Gamma_{\uparrow,m}(\Psi_k) \times \nonumber \\ &&\left\lvert  \bra{0}  \ e^{-\frac{1}{2}\sum_n [\Gamma_{\downarrow,n}(\Psi_k) \ket{1}\bra{1}+\Gamma_{\uparrow,n}(\Psi_k) \ket{0}\bra{0}]t} \ket{\psi} \right\rvert^2 \nonumber \\ &=&|\alpha_0|^2,
 \end{eqnarray}
giving total energy $E$. Thus, both measurement schemes produce equivalent energy distributions.

In the case that the calorimeter reaches the macrostate immediately after a jump, the calculation is similar with only two jump operators $\hat C_{\uparrow}(E)=\sqrt{\Gamma_{\uparrow}(E)} \hat{a}^\dagger$ and $\hat C_{\downarrow}(E)=\sqrt{\Gamma_{\downarrow,m}(E)} \hat{a}$.

\section{Fluctuation theorem for the quantum jump model with a finite-size environment} 

Let us consider an open quantum system coupled to a calorimeter through dissipative channels described by jump operators $\hat{D}_{m}=g_{m} \hat A_m \otimes \hat B_{m}$, where $ \hat A_m$ and $\hat B_{m}$ depend on the system and calorimeter degrees of freedom, respectively, and $g_{m}$ is the coupling strength. We also assume that the jump operator follow detailed balance such that for every $\hat D_{m}$ there is $\hat D_{n}$ such that $\hat A_n =\hat A_m^\dagger$ and $\hat B_{n} =\hat B_{m}^\dagger$. Let us first study the case, where the calorimeter is in a single microstate $\ket{\Psi_k}$ and it does not change between the jumps. In this case, we can use the calorimeter traced jump operators $\hat C_{m}=\sqrt{\Gamma_{m}(\Psi_k)} \hat{A}_m$ that are defined such that ${\rm Tr}_{s} \lbrace \hat C_{m} \hat \sigma_s \hat C^\dagger_{m}   \rbrace = {\rm Tr}_{s+c} \lbrace \hat D_{m} \hat \sigma_s \otimes \ket{\Psi_k} \bra{\Psi_k}  \hat D^\dagger_{m}   \rbrace$, where $\hat \sigma_s$ is the matrix form of the system state, $\rm Tr_{s}$ denotes trace over the system degrees of freedom and $\rm Tr_{s+c}$ denotes trace over both the system and calorimeter degrees of freedom. Due to the system-calorimeter energy change, the energy of the calorimeter evolves stochastically. The probability to traverse a single $N$-jump QJ trajectory is given by
\begin{eqnarray}
&& P_{QJ}[i, f, \Psi_0, \Psi_N, \{\hat{C}_{m_k}\}_{k=1}^N, \{t_k\}_{k=1}^N]  \\ 
&&= P[i,\Psi_0] \left[ \prod_{k=1}^N p^0(t_k,t_{k-1}) p_{m_k}(t_k)  \right]  p^0(\tau, t_N) P_f[f, \Psi_N], \nonumber
\label{Eq:PrF}
\end{eqnarray}
where the protocol starts at time $t_0=0$ and ends at time $\tau$, $P[i,\Psi_0]$ is the probability to start with the system state $\ket{i}$ and the calorimeter state $\ket{\Psi_0}$, $p_{m_k}(t_k)$ is the probability for a jump to occur along the $m_k$:th channel during $[t_k,t_k+\delta t]$, and $p^0(t_{k+1},t_k)$ is the probability of no jump during the time-interval $[t_k,t_{k+1}]$ and $P_f[f, \Psi_N]$ is the probability to measure the system state $\ket{f}$ and the calorimeter state $\ket{\Psi_N}$ at the end of the protocol. As the jump probabilities can be calculated simply using the calorimeter traced jump operators, we can use the results derived for time-dependent jump operators \cite{Suomela2015a}, yielding
\begin{eqnarray}
&&P_{QJ}[i, f, \Psi_0, \Psi_N, \{\hat{C}_{m_k}\}_{k=1}^N, \{t_k\}_{k=1}^N] \nonumber  \\
&&=  (\delta t)^N P[i,\Psi_0] \left[\prod_{k=1}^N \Gamma_{m_k}(\Psi_{k-1}) \right] \times \\ &&\left\lvert \langle f | \hat{U}_{\mathrm{eff}}(\tau,t_N) \left[ \prod_{k=1}^N \hat{A}_{m_{N+1-k}}\hat{U}_{\mathrm{eff}}(t_{N+1-k},t_{N-k})\right] |i \rangle \right\rvert^2, \nonumber
\label{Eq:Pr1}
\end{eqnarray}

where $\ket{\Psi_k}$ is the calorimeter state after $k$:th jump and the no-jump evolution is given by
\begin{equation}
\hat{U}_{\mathrm{eff}}(t_{k+1},t_{k})= \mathcal{T} e^{ -\frac{i}{\hbar} \left[ \int_{t_{k}}^{t_{k+1}} \hat{H}_s(t)-i \frac{\hbar}{2} \sum_{i} \Gamma_i(\Psi_k)  \hat{A}_i^{\dagger}\hat{A}_i dt \right]},
\end{equation}
 where $\hat{H}_s(t)$ is the system Hamiltonian and $\mathcal{T}$ is the time-ordering operator.

We can formulate a time-reversed counterpart for the forward trajectory of Eq. \eqref{Eq:PrF}. In the time-reversed trajectory, we measure system state $\ket{f}$ and calorimeter state $\ket{\Psi_N}$ at the beginning ($\bar{t} = 0$) and states $\ket{i}$ and $\ket{\Psi_0}$ at the end ($\bar{t}=\tau$). In the time-reversed trajectory, all the jumps are reversed and they happen in reverse order, i.e., a jump caused by $\hat C_{\bar{m}_k}$ occurs at time  $\bar{t} = t_{f}-t_{k}$, where the index $\bar{m}_k$ is related to the forward index $m_k$ such that $\hat A_{\bar{m}_k}=\hat{A}^\dagger_{m_k}$ and $\hat B_{\bar{m}_k}=\hat{B}^\dagger_{m_k}$. By demanding that the time-reversed no-jump evolution between jumps is given by $\hat{U}_{\mathrm{eff}}^\dagger(t_{i+1},t_{i})$, the probability for the reverse QJ trajectory can then be written as 
\begin{equation}
\begin{split}
&\bar{P}_{QJ}[f, i, \Psi_N, \Psi_0, \{\hat{C}_{\bar{m}_k}\}_{k=1}^N, \{\bar{t}_k\}_{k=1}^N] \\
&=  (\delta t)^N \bar{P}[f,\Psi_N] \left[ \prod_{k=1}^N {\Gamma}_{\bar{m}_k}(\Psi_{k}) \right] \times \\ &\left\lvert  \langle i | \left[ \prod_{k=1}^N \hat{U}^\dagger_{\mathrm{eff}}(t_{k},t_{k-1}) \hat{A}^\dagger_{m_{k}}  \right]  \hat{U}^\dagger_{\mathrm{eff}}(\tau,t_{N})| f \rangle \right\rvert^2,
\label{Eq:Pr1}
\end{split}
\end{equation}
where $\bar{P}[f,\Psi_N]$ is the probability to start a reversed trajectory with the system state $\ket{f}$ and the calorimeter state $\ket{\Psi_N}$. The ratio of the forward and reversed trajectory probabilities is of the form:
\begin{equation}
\begin{split}
&\Delta S_{T}[i, f, \Psi_0, \Psi_N, \{\hat{C}_{m_k}\}_{k=1}^N, \{t_k\}_{k=1}^N] \\ &= \ln \left[ \frac{P_{QJ}[i, f, \Psi_0, \Psi_N, \{\hat{C}_{m_k}\}_{k=1}^N, \{t_k\}_{k=1}^N]}{\bar{P}_{QJ}[f, i, \Psi_N, \Psi_0, \{\hat{C}_{\bar{m}_k}\}_{k=1}^N, \{\bar{t}_k\}_{k=1}^N] } \right] \\
&=\ln \left\lbrace  \frac{P[i,\Psi_0]}{\bar{P}[f,\Psi_N]}\right\rbrace+\ln \left[ \prod_{k=1}^{N} \frac{\Gamma_{m_k}(\Psi_{k-1})}{\Gamma_{\bar{m}_k}(\Psi_{k})}  \right].
\end{split}
\label{Eq:R}
\end{equation}
We denote this term as the total entropy production of the model. As the reversed trajectories' probabilities sum up to unity, it can be straightforwardly shown that
\begin{equation}
\langle e^{-\Delta S_{T}} \rangle = 1, \label{Eq:EFT}
\end{equation}
where the average is over all the forward trajectories.

As discussed in the main text, the transition rates satisfy the condition $\Gamma_{\bar{m}_k}(\Psi_{k})=\Gamma_{m_k}(\Psi_{k-1})$ when the calorimeter is assumed to be in a single microstate. As a consequence, the second term of Eq. \eqref{Eq:R} is zero and the entropy production depends only on the forward and reversed trajectory initial probability distributions. If the initial probability distribution of the forward trajectory follows canonical ensemble, the probability distribution of the reversed trajectories can be chosen to follow a canonical ensemble with the same temperature. In this case, the total entropy production becomes equivalent with the energy difference between the final and initial state of the total system. By defining the work of a single trajectory as the energy difference between the final and initial  state of the total system, Eq. \eqref{Eq:EFT} gives the Jarzynski equality.

If the relaxation rate inside the calorimeter is the fastest time scale, then the calorimeter does not stay in a single microstate between jumps but shifts quickly between the microstates  $\ket{\Psi_k}$ that correspond to the same energy. We can still derive a theorem similar to Eq. \eqref{Eq:R} with $\Psi_k$ denoting calorimeter energies instead of states. The transition rates are then given by Eqs. \eqref{ETR1} and \eqref{ETR2} of the main text.  In this case, the product of the transition rate ratio gives $N(E_0)/N(E_N)$ in Eq. \eqref{Eq:R}, where  $E_0$ and $E_N$ are the initial and final energies of the calorimeter, respectively, and $N(E)$ denotes the number of microstates corresponding energy $E$. If both the forward and backward process start from the canonical ensemble, i.e., $ P[i,E_0]=N(E_0) e^{-\beta ( \hbar \omega_i+E_0)}/{Z}$ and $\bar{P}[f,E_N]=N(E_N) e^{-\beta ( \hbar \omega_f+E_N)}/{Z^\prime}$, we get again the Jarzynski equality with $\Delta F = - (1/\beta) \ln (Z^\prime/Z)$. Here, $\hbar \omega_i$ and $\hbar \omega_f$ denote the energies of  system states $\ket{i}$ and $\ket{f}$, respectively.

\section{Detailed balance condition for the transition rates}

When the energy fluctuations are small in the canonical ensemble, they are often approximated by a gaussian distribution around the average energy $E_0$. If the heat capacity $C$ is constant, these energy fluctuations are often expressed as the effective temperature fluctuations $T=E/C$ in mesoscopic systems \cite{Landau1980statistical,PhysRevLett.102.130605,jsm2013/P02033}. We will now show that the assumption of a constant heat capacity can lead to violation of the main text's Eq. \eqref{eq:DBC2} when temperature fluctuations $T$ are used together with the standard temperature-dependent transition rates 
 \begin{eqnarray}
\Gamma_{\downarrow}(\hbar \omega_0 ,T(E))= |g|^2/(1\pm e^{-\hbar \omega_0/(k_BT)}), \label{eq:TR1}\\
\Gamma_{\uparrow}(\hbar \omega_0 ,T(E))= |g|^2/(e^{\hbar \omega_0/(k_BT)} \pm 1), \label{eq:TR2}
\end{eqnarray}
where $g$ is the coupling strength, $\hbar \omega_0$ is the energy gap of the qubit, $T(E)$ is the effective temperature corresponding to the calorimeter energy $E$ and $+$ $(-)$ is used in the case fermionic (bosonic) transition rates. If a jump up occurs in the qubit, the energy of the calorimeter decreases by $\hbar \omega_0$ and consequently the effective temperature decreases by $\hbar \omega_0/C$. 

For Gaussian energy fluctuations, the left-hand side of Eq. \eqref{eq:DBC2} simplifies to
\begin{eqnarray}
\frac{N(E)}{N(E-\hbar \omega_0)}=e^{\hbar \omega_0\left \lbrace { \beta +\left[\hbar \omega_0 - 2(E-E_0) \right] /(2\sigma^2)}\right \rbrace}, \label{eq:A2Ns}
\end{eqnarray}
where $N(E)$ and $N(E-\hbar \omega_0)$ are the number of microstates corresponding to calorimeter energies $E$ and $E-\hbar \omega_0$, respectively, $\beta$ is the inverse temperature of the ideal bath that is used to thermalized both the qubit and the calorimeter and $\sigma^2=C k_B^{-1} \beta^{-2} $ is the variance of the gaussian calorimeter energy distribution. However, if now the transition rates of Eqs. \eqref{eq:TR1}-\eqref{eq:TR2} are used, the right-hand side of Eq. \eqref{eq:DBC2} does not agree with Eq. \eqref{eq:A2Ns} and thus the detailed balance condition is not satisfied. This explains the apparent violations of Jarzynski equality in Ref. \cite{jsm2013/P02033}, where similar type of transition rates were used with gaussian effective temperature fluctuations and a constant heat capacity.

\bibliography{suomela2,localbib}

\end{document}